\documentclass[conference]{IEEEtran}
\IEEEoverridecommandlockouts
\usepackage{cite}
\usepackage{amsmath,amssymb,amsfonts}
\usepackage{graphicx}
\usepackage{textcomp}
\usepackage{xcolor}
\usepackage{algorithm}
\usepackage{algpseudocode}

\usepackage[left=1.62cm,right=1.62cm,top=1.9cm]{geometry}

\begin{document}

\title{Deep Q-Network for Optimizing NOMA-Aided Resource Allocation in Smart Factories with URLLC Constraints
}

\author{\IEEEauthorblockN{Shi Gengtian, Jiang Liu, Shigeru Shimamoto}
\IEEEauthorblockA{\textit{Graduate School of Fundamental Science and Engineering} \\
\textit{Waseda University, Tokyo, Japan 169-8050}\\
Email: shigengtian@akane.waseda.jp, jiang@waseda.jp, shima@waseda.jp}
}
\maketitle

\begin{abstract}
This paper presents a Deep Q-Network (DQN)-based algorithm for NOMA-aided resource allocation in smart factories, addressing the stringent requirements of Ultra-Reliable Low-Latency Communication (URLLC). The proposed algorithm dynamically allocates sub-channels and optimizes power levels to maximize throughput while meeting strict latency constraints. By incorporating a tunable parameter  $\lambda$, the algorithm balances the trade-off between throughput and latency, making it suitable for various devices, including robots, sensors, and controllers, each with distinct communication needs. Simulation results show that robots achieve higher throughput, while sensors and controllers meet the low-latency requirements of URLLC, ensuring reliable communication for real-time industrial applications. 
\end{abstract}


\section{Introduction}
In the area of Industry 4.0, smart factories are revolutionizing manufacturing processes by leveraging advanced technologies such as the Internet of Things (IoT) \cite{wollschlaeger2017future}, artificial intelligence (AI), and wireless communication systems. These factories are characterized by interconnected devices, including robots, sensors, controllers, and other smart devices, which collaborate to optimize production efficiency, quality, and safety \cite{lee2015cyber} \cite{zhong2017intelligent}.

Wireless communication plays a pivotal role in smart factories, enabling real-time data exchange and control among diverse devices spread across the factory floor. However, the increasing density and diversity of devices pose significant challenges to traditional communication systems, including spectrum scarcity, interference management, and latency-sensitive applications \cite{sisinni2018industrial} \cite{cheng2018industrial}.

Non-Orthogonal Multiple Access (NOMA) emerges as a promising solution to address these challenges by enabling multiple users to share the same frequency band and time slot, thereby enhancing spectrum efficiency and accommodating a large number of connected devices \cite{ding2017application}. By exploiting power domain multiplexing and successive interference cancellation (SIC) techniques, NOMA offers a flexible algorithm for resource allocation and transmission scheduling in dense and dynamic communication environments \cite{dai2015non}.

In this context, optimizing NOMA-aided resource allocation is essential for maximizing system throughput, minimizing latency, and ensuring efficient use of wireless resources in smart factories. Reinforcement learning (RL) provides a promising approach by enabling autonomous decision-making in dynamic environments. Unlike traditional methods, RL adapts based on continuous interaction with the environment, making it particularly effective in the ever-changing landscape of smart factories.

This paper explores the application of RL to dynamically allocate sub-channels and power levels, aiming to optimize system throughput while meeting URLLC (Ultra-Reliable Low-Latency Communication) constraints. By employing a carefully designed reward function, the proposed RL-based algorithm learns effective resource allocation strategies through trial and error.

Our experimental results demonstrate that the proposed approach significantly improves communication performance and resource utilization efficiency in NOMA-enabled smart factories. These findings advance the state-of-the-art in wireless communication systems for Industry 4.0, paving the way for more adaptive, intelligent factory automation systems.

\subsection{Related Work}
The vision of smart factories enabled by the Industrial Internet of Things (IIoT) has driven significant research into reliable low-latency wireless communication technologies. Early WiFi and cellular network generations lacked the stringent quality-of-service (QoS) requirements for mission-critical industrial automation \cite{sisinni2018industrial}. Dedicated protocols like WirelessHART \cite{song2008wirelesshart} and ISA100.11a \cite{salvadori2009monitoring} offered improved reliability but still suffered from substantial latency limitations.
The advent of 5G's Ultra-Reliable Low Latency Communication (URLLC) service opened new possibilities by defining strict targets of less than 1ms latency and 
{\(10^{-5}\)}
packet loss rates \cite{berardinelli2018beyond}. This has catalyzed substantial research on leveraging URLLC for real-time industrial control and monitoring \cite{aijaz2020private}. However, efficiently allocating limited time/frequency resources to satisfy diverse URLLC traffic demands in dense IIoT environments remains an open challenge \cite{ali2021urllc}.
Prior work has applied reinforcement learning (RL) to general wireless resource allocation problems \cite{gengtian2020power} , but these techniques often rely on oversimplified network models and heuristic reward functions which may not translate well to dynamic smart factory settings. Some recent studies have begun exploring RL specifically for URLLC resource management \cite{she2019ultra}, but remain limited to basic simulation environments and struggle to satisfy the extreme QoS constraints.
% In contrast, our approach leverages a high-fidelity digital twin simulation of a real factory floor to capture intricate spatio-temporal traffic patterns and system dynamics.

Additionally, we meticulously designed the reinforcement learning environment with a tailored state space, action set, and reward function aimed at directly optimizing critical performance metrics such as packet delivery ratios and age-of-information delays. This deliberate design empowers our method to learn sophisticated resource allocation strategies that closely align with operational requirements.

Existing solutions have not adequately tackled the problem of dynamic resource allocation for diverse URLLC flows within smart factories while considering practical real-world factors like user mobility, obstructions, and unpredictable traffic bursts. Our proposed RL-based strategy using URLLC technology aims to bridge this gap and enable mission-critical industrial automation over 5G networks with sufficient reliability and low latency.

\subsection{Contribution}
This paper makes the following key contributions to the field of intelligent manufacturing systems and wireless communication:

\begin{itemize}
    \item \textbf{DQN-Based Resource Allocation}: We propose a novel Deep Q-Network (DQN)-based algorithm for optimizing NOMA-aided resource allocation, tailored to meet the diverse needs of devices in smart factory environments.
    \item \textbf{Throughput-Latency Trade-off}: The introduction of a tunable parameter \( \lambda \) enables dynamic balancing between throughput and latency, allowing the system to meet the distinct communication requirements of robots, sensors, and controllers in URLLC scenarios.
    \item \textbf{URLLC Considerations}: Our approach ensures that latency-sensitive devices, such as sensors and controllers, meet ultra-reliable low-latency requirements, while high-throughput devices like robots maintain efficient resource utilization.
    \item \textbf{Simulation and Performance Evaluation}: We conduct extensive simulations to demonstrate the effectiveness of the proposed algorithm in optimizing throughput and latency across various scenarios, contributing to future advancements in industrial automation.
\end{itemize}

By addressing these aspects, our work contributes to the advancement of intelligent manufacturing systems, enhancing their efficiency, reliability, and overall performance through optimized wireless communication strategies.

\section{SYSTEM MODEL}
\begin{figure}[htbp] 
\centering
    {\includegraphics[scale=0.35]{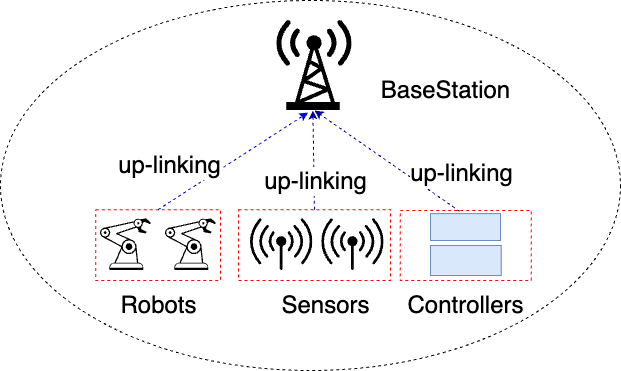}}
% \caption{System Model }
\label{fig:system_model}
\end{figure}

In the communication environment of smart factories, there is a base station (BS) located at the center of the factory. As illustrated in Fig. \ref{fig:system_model}, The BS communicates with $N_s$ user devices (including robots, sensors, and controllers) through $N_s$ orthogonal sub-channels. All user devices and the base station are assumed to be equipped with a single antenna.

\subsection{System Components}
The system consists of $N_s$ orthogonal sub-channels, denoted by $\mathcal{Y}_s = \{ s_1, s_2, \ldots, s_{N_s} \}$, and $N_u$ user devices, denoted by $\{ u_1, u_2, \ldots, u_{N_u} \}$, categorized into three types: robots, sensors, and controllers. The data block size for each device is the same, denoted by $m_k \in \mathcal{Y}_m$ ($k \in [1, N_u]$), where each data block consists of $D$ bits. The transmission of a data block on each sub-channel must be completed within $M$ seconds per unit bandwidth.

% \vspace{4pt}  % Adjust top space if needed
% \begin{figure*}[!t]
%     \centering
%     \includegraphics[width=\textwidth, height=0.20\textheight, trim=0 15 0 5, clip]{dqn_3.jpg}
%     \caption{Deep Q-Network (DQN) Architecture}
%     \label{fig:dqn}
% \end{figure*}

% \vspace{5pt}  % Add subtle spacing before the figure

\subsection{Device Communication Requirements}
Robots are primarily responsible for mobility and task operation, requiring high bandwidth and low latency. Sensors are used for environmental monitoring and status detection, typically transmitting small data packets but requiring low latency and high reliability. Controllers are responsible for controlling and coordinating the work of robots and sensors, requiring high bandwidth and low latency.

\subsection{Channel Allocation and Power Allocation}
Each user device can be assigned one or more sub-channels. The set of users connected to the base station via sub-channel $s_j$ (i.e., NOMA clusters) is denoted by $\mathcal{Y}_j^u = \{ u_1, u_2, \ldots, u_{N_j^u} \}$, where $N_j^u$ is the number of users connected to the base station via sub-channel $s_j$, and $\sum_{j=1}^{N_s} N_j^u = N_u$. The power allocation for device $i$ is denoted by $p_i$, and for simplicity, it can be assumed that each device can choose discrete power levels.

\vspace{5pt}
\begin{figure*}[htbp] 
    \centerline{\includegraphics[scale=0.4]{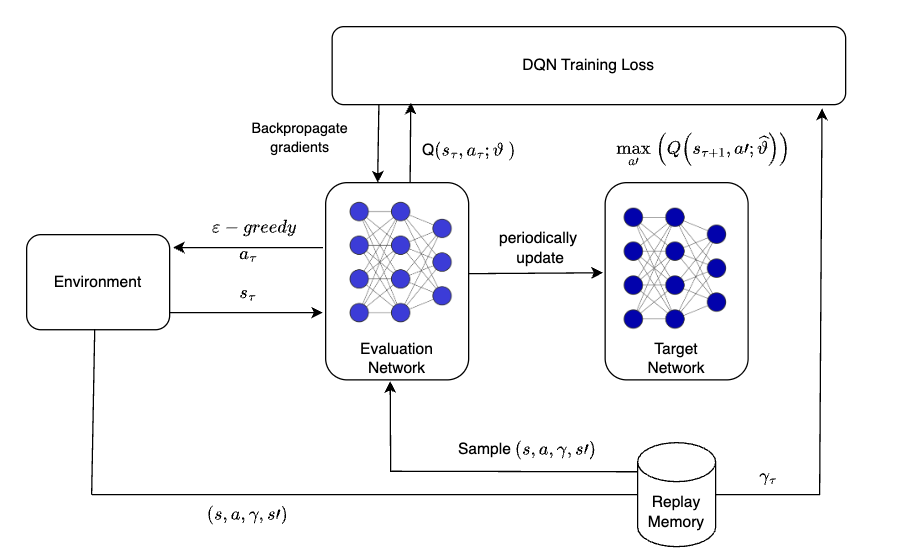}}
    \caption{Deep Q-Network (DQN) Architecture}
    \label{fig:dqn}
\end{figure*}

% \subsection{Channel Gain and Path Loss}
\subsection{Channel Gain and Path Loss}
Channel gain \( h_i \) and path loss \( PL(d_i) \) directly affect communication quality. The channel gain is given by:  
\begin{equation}
    h_i = \frac{g_i}{d_i^n}
\end{equation}
where \( g_i \) is the small-scale fading coefficient, \( d_i \) is the distance, and \( n \) is the path loss exponent.

Path loss is an essential metric in wireless communication, representing signal attenuation over distance where \( PL(d_0) \) is the reference path loss, and \( X\sigma \) accounts for shadow fading.
Where $d_i$ is the distance between device $i$ and the base station, $PL(d_0)$ is the path loss at the reference distance $d_0$, $n$ is the path loss exponent, and $X \sigma$ represents the Gaussian random variable for shadow fading.
\begin{equation}
    PL(d_i) = PL(d_0) + 10n \log_{10} \left( \frac{d_i}{d_0} \right) + X\sigma
\end{equation}
% where \( PL(d_0) \) is the reference path loss, and \( X\sigma \) accounts for shadow fading.
% Where $d_i$ is the distance between device $i$ and the base station, $PL(d_0)$ is the path loss at the reference distance $d_0$, $n$ is the path loss exponent, and $X \sigma$ represents the Gaussian random variable for shadow fading.

\subsection{Throughput Calculation}
The throughput of each device can be calculated by the following formula:

\begin{equation}
T_i = \log_2 \left(1 + \frac{h_i p_i}{\sum_{j \neq i} h_j p_j + \sigma^2}\right)
\end{equation}

Where $\sigma^2$ represents the noise power.

\subsection{Problem Statement}
The objective is to optimize the wireless communication resource allocation in smart factories by maximizing the total throughput of the system while minimizing communication latency. The optimization problem can be formulated as maximizing the total throughput subject to latency, power, and channel constraints. The latency constraint requires that the transmission of a data block on each sub-channel must be completed within $M$ seconds per unit bandwidth. The power constraint dictates that the transmission power of each device must be within its maximum power $P_{\text{max}}$. The channel constraint specifies that each sub-channel can serve multiple users simultaneously, but each user can occupy only one sub-channel.

\section{Reinforcement Learning in NOMA-Aided Resource Allocation}
In the context of optimizing NOMA-aided resource allocation in smart factories, reinforcement learning (RL) offers a promising approach. A single-agent RL model can dynamically allocate sub-channels and optimize power levels for user devices, aiming to maximize throughput while minimizing latency.
The problem is modeled as a Markov Decision Process (MDP), where the state space includes information such as channel conditions and system parameters. The state at time \( t \) can be represented as:
\begin{equation}
s_t = \{h_t, p_t\}
\end{equation}
where \( h_t \) denotes the channel condition, and \( p_t \) represents the power allocation at time \( t \).

The action space involves resource allocation decisions, specifically selecting sub-channels and assigning power levels. The action at time \( t \) is defined as:
\begin{equation}
a_t = (c_t, p_t)
\end{equation}
where \( c_t \) is the sub-channel selected, and \( p_t \) is the corresponding power level.

The reward function guides the learning process by providing feedback on the quality of actions. It encourages actions that maximize throughput and penalizes those that cause high latency or inefficient resource use. The reward at time \( t \) is formulated as:
\begin{equation}
r_t = \log_2\left(1 + \frac{h_t p_t}{\sigma^2 + \sum_{j \neq i} h_j p_j}\right) - \lambda L_t
\end{equation}
where \( \sigma^2 \) is the noise power, \( L_t \) is the latency, and \( \lambda \) is a factor balancing throughput and latency.

% \section{Deep Q-Network (DQN)}
As shown in Fig. \ref{fig:dqn}, the DQN architecture consists of an agent interacting with the environment by selecting actions through an epsilon-greedy policy. The evaluation network computes the Q-values for the given state-action pairs, while the target network, updated periodically by copying the weights of the evaluation network, provides stability in training. Transitions, including the state, action, reward, and next state, are stored in replay memory to remove temporal correlations in the training data. The loss is computed as the difference between the Q-values predicted by the evaluation network and the target values, with the evaluation network updated via gradient descent to minimize this loss.

The Q-value function is updated using the following rule:

\begin{equation}
\begin{split}
Q(s_t, a_t) \leftarrow Q(s_t, a_t) + \alpha \Big[ r_t + \gamma \max_{a'} Q(s_{t+1}, a') \\
- Q(s_t, a_t) \Big]
\end{split}
\end{equation}

where \( \alpha \) is the learning rate, \( \gamma \) is the discount factor and \( r_t \) is the reward at time \( t \).

\begin{algorithm}[H]
\caption{Deep Q-Network (DQN) Training}
\begin{algorithmic}[1]
\State Initialize Q-network \( Q \) and target network \( Q' \)
\State Initialize replay buffer \( D \)
\For{each episode}
    \State Initialize environment and receive initial state \( s_0 \)
    \For{each time step \( t \)}
        \State With probability \( \epsilon \), select a random action \( a_t \)
        \State Otherwise, select \( a_t = \arg\max_a Q(s_t, a) \)
        \State Execute action \( a_t \) and observe reward \( r_t \) and next state \( s_{t+1} \)
        \State Store transition \( (s_t, a_t, r_t, s_{t+1}) \) in replay buffer \( D \)
        \State Sample a random mini-batch of transitions from \( D \)
        \State Compute the target for each transition:
        \[
        y_t = r_t + \gamma \max_{a'} Q'(s_{t+1}, a')
        \]
        \State Update Q-network \( Q \) by minimizing the loss:
        \[
        \mathcal{L}(\theta) = \mathbb{E}[(y_t - Q(s_t, a_t))^2]
        \]
        \State Periodically update target network: \( Q' \leftarrow Q \)
        \State Set \( s_t = s_{t+1} \)
    \EndFor
\EndFor
\end{algorithmic}
\label{alg:dqn_training}
\end{algorithm}

% \begin{algorithm}[H]
% \caption{Deep Q-Network (DQN) Inference}
% \begin{algorithmic}[1]
% \State Initialize environment
% \State Load trained Q-network \( Q \)
% \For{each episode}
%     \State Initialize environment and receive initial state \( s_0 \)
%     \For{each time step \( t \)}
%         \State Select action \( a_t = \arg\max_a Q(s_t, a) \)
%         \State Execute action \( a_t \) and observe next state \( s_{t+1} \) and reward \( r_t \)
%         \State Set \( s_t = s_{t+1} \)
%     \EndFor
%     \State Log the total rewards for the episode
% \EndFor
% \end{algorithmic}
% \end{algorithm}

% In the training phase, the agent interacts with the environment, selecting actions based on an exploration-exploitation strategy. The transitions are stored in the experience replay buffer, and mini-batches are sampled to update the Q-network. Th
% e target network is periodically updated to stabilize the training process.

% In the inference phase, the trained Q-network is used to select actions in real-time, based on the learned policy, without further updates to the network.

As shown in Algorithm~\ref{alg:dqn_training}, the agent interacts with the environment, selecting actions based on an exploration-exploitation strategy to balance discovering new actions and maximizing rewards using the current policy. Each interaction generates a transition tuple consisting of the current state, selected action, received reward, and next state, which is stored in the experience replay buffer. Mini-batches of transitions are randomly sampled to update the Q-network, helping to break correlations between consecutive experiences and improve stability. A separate target network is periodically updated to further stabilize training and enhance convergence.

% \noindent
In the inference phase, the trained Q-network is deployed for real-time decision-making, selecting optimal actions based on the learned policy without further updates. Since exploration is no longer required, the agent fully exploits its learned knowledge to maximize performance. The computational efficiency of inference is crucial, especially in real-time applications where fast response times are essential.

\section{SIMULATION AND ANALYSIS}

\begin{table}[ht]
\centering
\caption{Simulation Parameters}
\begin{tabular}{||l||c||}
\hline
\textbf{Parameter} & \textbf{Value} \\
\hline
Number of Episodes & 1000 \\
\hline
Max Timesteps per Episode & 200 \\
\hline
Sub Channel Number & 10 \\
\hline
White Noise Power ($\sigma^2$) & 0.1 \\
\hline
Path Loss Parameter ($n$) & 2 \\
\hline
Number of Robots & 5 \\
\hline
Number of Sensors & 10 \\
\hline
Number of Controllers & 10 \\
\hline
Discount Factor ($\gamma$) & 0.99 \\
\hline
Batch Size & 64 \\
\hline
Number of Neurons (Hidden Layers) & 128 \\
\hline
Memory Size & 2000 \\
\hline
Bandwidth & 200 MHz \\
\hline
Noise Power & 1e-6 \\
\hline
Data Size (Robot) & 1500 bytes \\
\hline
Data Size (Sensor) & 1024 bytes \\
\hline
Data Size (Controller) & 512 bytes \\
\hline
\end{tabular}
\label{tab:simulation_parameters}
\end{table}

To evaluate the performance of our proposed RL-based resource allocation algorithm in a smart factory environment, we conducted extensive simulations using specific parameters, as shown in Table \ref{tab:simulation_parameters}. We trained the RL agent over 1000 episodes, with each episode comprising a maximum of 200 timesteps. The communication system utilized 10 sub-channels, allowing multiple devices to share the medium via NOMA. The power of the additive white Gaussian noise (AWGN) in the environment was set to 0.1, influencing the signal quality. A path loss parameter of 2 was used in the path loss model to simulate signal attenuation over distance, affecting the channel gain between the base station and user devices.

The simulation environment included 5 robots requiring high bandwidth and low latency, 10 sensors for environmental monitoring and status detection with low latency and high-reliability requirements, and 10 controllers responsible for coordinating the activities of robots and sensors, necessitating high bandwidth and low latency. The communication system was configured with a total bandwidth of 200 MHz, allowing sufficient capacity for data transmission. The noise power in the environment was set to 1e-6, simulating background interference that could affect communication quality. The data size for communication was defined as 1500 bytes for robots, 1024 bytes for sensors, and 512 bytes for controllers, representing the typical packet sizes for each device type in the simulation. The learning rate for the neural network training in the DQN algorithms was set to 0.001, balancing the convergence speed and stability of learning. A discount factor of 0.99 was used to emphasize future rewards in the DQN algorithm, while a batch size of 64 was chosen to determine the number of samples per training iteration, impacting the accuracy and stability of gradient estimation. The neural network used in the DQN algorithms had 128 neurons in its hidden layers, providing sufficient capacity to learn complex representations. Additionally, the memory size for experience replay was set to 2000, ensuring a large buffer for storing transitions and improving the stability of training.

\begin{figure}[htbp]
    \centerline{\includegraphics[scale=0.23]{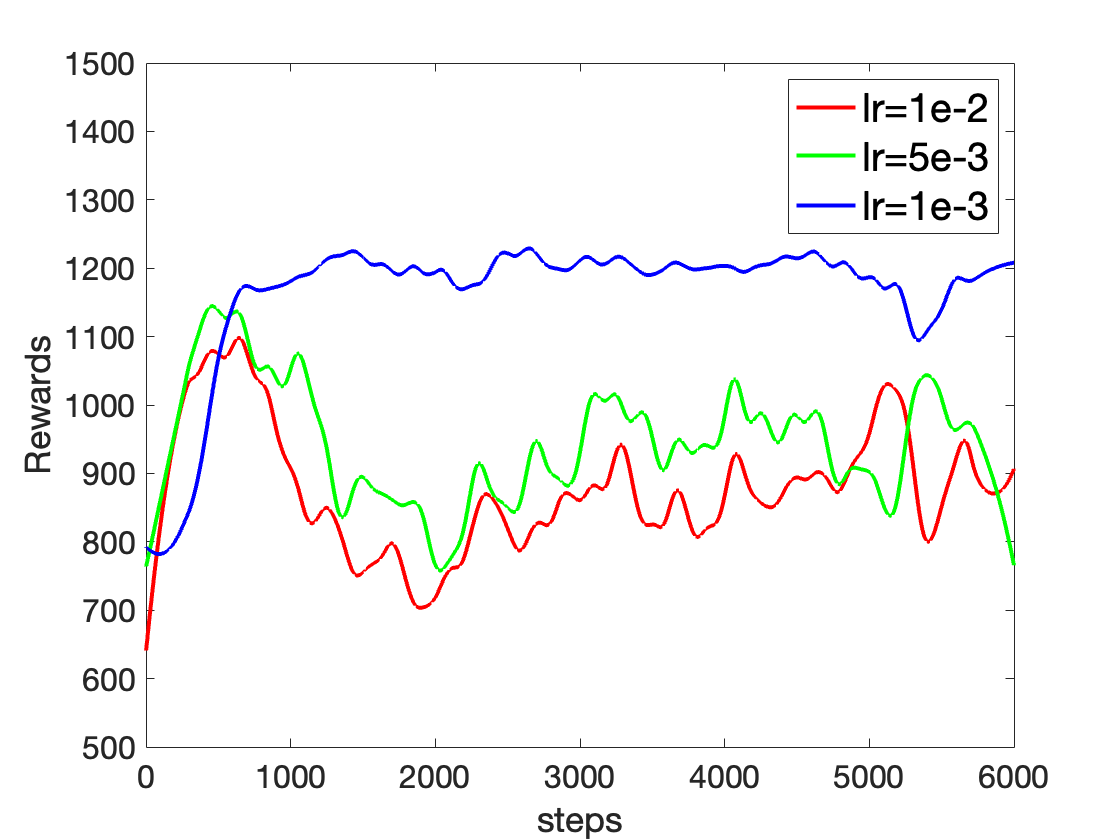}}
    \caption{Reward vs. Steps for different learning rates (lr).}
    \label{fig:reward_vs_steps}
\end{figure}

\begin{figure}[htbp]
    \centerline{\includegraphics[scale=0.45]{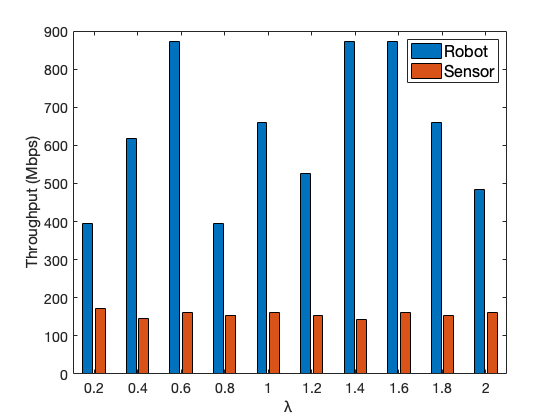}}
    \caption{Throughput (Mbps) for Robots and Sensors across different \( \lambda \).}
    \label{fig:throughput_vs_lambda}
\end{figure}

\begin{figure}[htbp]
    \centerline{\includegraphics[scale=0.45]{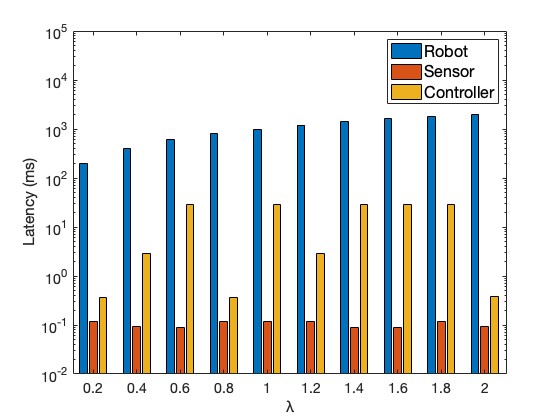}}
    \caption{Latency (ms) for Robots, Sensors, and Controllers at different \( \lambda \).}
    \label{fig:latency_vs_lambda}
\end{figure}

\section{Experimental Results}
We evaluated the impact of different learning rates on the performance of our RL-based resource allocation strategy. Fig. \ref{fig:reward_vs_steps} shows the total rewards over 6000 steps for learning rates  1e-2 ,  5e-3 , and  1e-3. With  lr = 1e-3  (blue curve), the RL agent achieves the highest and most stable rewards, converging smoothly around 1200 with minimal fluctuations. This suggests that  1e-3  is the optimal learning rate for this task.In contrast,  lr = 5e-3  (green curve) leads to moderate performance, with rewards fluctuating more and stabilizing around 1000. The learning rate  1e-2  (red curve) results in the worst performance, with significant oscillations and lower rewards, indicating unstable learning.

Fig. \ref{fig:throughput_vs_lambda} illustrates the throughput (Mbps) for robots and sensors across varying values of  $\lambda$ , which balances the trade-off between throughput and latency. Robots exhibit significantly higher throughput compared to sensors, with peak values reaching around 900 Mbps. This suggests that robots, which typically require more bandwidth and have stricter low-latency requirements, are prioritized in the resource allocation process. In contrast, Sensors, on the other hand, maintain a lower throughput throughout the simulations, fluctuating between 100 and 200 Mbps. This reflects their lower bandwidth demands compared to robots. While the throughput for sensors remains relatively stable, it still varies with changes in  $\lambda$ , suggesting that sensor resource allocation is influenced by the same trade-off factor. The variation in throughput for both robots and sensors as  $\lambda$  changes indicates that  $\lambda$  plays an important role in adjusting the resource allocation between devices with different communication requirements. Overall, robots consistently benefit more from higher throughput than sensors across all values of  $\lambda$. Robot throughput variation depends on channel conditions, interference, and DQN stability. If $\lambda$ increases but resources are limited, throughput improvement may be restricted.

% \section{Latency Results with URLLC Considerations}
Fig. \ref{fig:latency_vs_lambda} illustrates the latency (ms) for robots, sensors, and controllers across varying $\lambda$, which balances throughput and latency in URLLC applications. Robots exhibit the highest latency (often $>$100 ms) due to high bandwidth demands, while sensors maintain low latency ($<$10 ms), meeting URLLC requirements. Controllers fall in between (10-100 ms). As $\lambda$ increases, robot latency slightly decreases, while sensors and controllers remain stable. These findings emphasize the need for optimized resource allocation to ensure low latency and high reliability in smart factories.

\section{CONCLUSION AND FUTURE WORK}
In this research, we developed a DQN-based algorithm for NOMA-aided resource allocation in smart factories, with a focus on meeting URLLC constraints. The proposed approach demonstrated its effectiveness in balancing the trade-off between throughput and latency, ensuring that robots, with their higher bandwidth demands, achieved greater throughput, while sensors and controllers maintained the low latency required by URLLC. The inclusion of the  $\lambda$  parameter allowed for flexible adjustments between latency and throughput, making the algorithm suitable for diverse industrial environments.

For future work, exploring multi-agent reinforcement learning (MARL) can enable decentralized learning, optimizing each device’s policy individually. Integrating advanced RL methods like PPO or actor-critic can enhance training stability and performance. Expanding the algorithm to handle heterogeneous devices, mobility, fading, and interference will improve applicability in industrial IoT. Lastly, developing energy-efficient strategies will be key to balancing power consumption and communication performance in smart factories.

\section{Acknowledgments}
This study was supported by Waseda Research Institute for Science and Engineering project research.

\bibliographystyle{IEEEtran}
\bibliography{ref}  % 这里对应 references.bib

\begin{thebibliography}{10}
\providecommand{\url}[1]{#1}
\csname url@samestyle\endcsname
\providecommand{\newblock}{\relax}
\providecommand{\bibinfo}[2]{#2}
\providecommand{\BIBentrySTDinterwordspacing}{\spaceskip=0pt\relax}
\providecommand{\BIBentryALTinterwordstretchfactor}{4}
\providecommand{\BIBentryALTinterwordspacing}{\spaceskip=\fontdimen2\font plus
\BIBentryALTinterwordstretchfactor\fontdimen3\font minus \fontdimen4\font\relax}
\providecommand{\BIBforeignlanguage}[2]{{%
\expandafter\ifx\csname l@#1\endcsname\relax
\typeout{** WARNING: IEEEtran.bst: No hyphenation pattern has been}%
\typeout{** loaded for the language `#1'. Using the pattern for}%
\typeout{** the default language instead.}%
\else
\language=\csname l@#1\endcsname
\fi
#2}}
\providecommand{\BIBdecl}{\relax}
\BIBdecl

\bibitem{wollschlaeger2017future}
M.~Wollschlaeger, T.~Sauter, and J.~Jasperneite, ``The future of industrial communication: Automation networks in the era of the internet of things and industry 4.0,'' \emph{IEEE industrial electronics magazine}, vol.~11, no.~1, pp. 17--27, 2017.

\bibitem{lee2015cyber}
J.~Lee, B.~Bagheri, and H.-A. Kao, ``A cyber-physical systems architecture for industry 4.0-based manufacturing systems,'' \emph{Manufacturing letters}, vol.~3, pp. 18--23, 2015.

\bibitem{zhong2017intelligent}
R.~Y. Zhong, X.~Xu, E.~Klotz, and S.~T. Newman, ``Intelligent manufacturing in the context of industry 4.0: a review,'' \emph{Engineering}, vol.~3, no.~5, pp. 616--630, 2017.

\bibitem{sisinni2018industrial}
E.~Sisinni, A.~Saifullah, S.~Han, U.~Jennehag, and M.~Gidlund, ``Industrial internet of things: Challenges, opportunities, and directions,'' \emph{IEEE transactions on industrial informatics}, vol.~14, no.~11, pp. 4724--4734, 2018.

\bibitem{cheng2018industrial}
J.~Cheng, W.~Chen, F.~Tao, and C.-L. Lin, ``Industrial iot in 5g environment towards smart manufacturing,'' \emph{Journal of Industrial Information Integration}, vol.~10, pp. 10--19, 2018.

\bibitem{ding2017application}
Z.~Ding, Y.~Liu, J.~Choi, Q.~Sun, M.~Elkashlan, I.~Chih-Lin, and H.~V. Poor, ``Application of non-orthogonal multiple access in lte and 5g networks,'' \emph{IEEE Communications Magazine}, vol.~55, no.~2, pp. 185--191, 2017.

\bibitem{dai2015non}
L.~Dai, B.~Wang, Y.~Yuan, S.~Han, I.~Chih-Lin, and Z.~Wang, ``Non-orthogonal multiple access for 5g: solutions, challenges, opportunities, and future research trends,'' \emph{IEEE Communications Magazine}, vol.~53, no.~9, pp. 74--81, 2015.

\bibitem{song2008wirelesshart}
J.~Song, S.~Han, A.~Mok, D.~Chen, M.~Lucas, M.~Nixon, and W.~Pratt, ``Wirelesshart: Applying wireless technology in real-time industrial process control,'' in \emph{2008 IEEE Real-Time and Embedded Technology and Applications Symposium}.\hskip 1em plus 0.5em minus 0.4em\relax IEEE, 2008, pp. 377--386.

\bibitem{salvadori2009monitoring}
F.~Salvadori, M.~de~Campos, P.~S. Sausen, R.~F. de~Camargo, C.~Gehrke, C.~Rech, M.~A. Spohn, and A.~C. Oliveira, ``Monitoring in industrial systems using wireless sensor network with dynamic power management,'' \emph{IEEE Transactions on Instrumentation and Measurement}, vol.~58, no.~9, pp. 3104--3111, 2009.

\bibitem{berardinelli2018beyond}
G.~Berardinelli, N.~H. Mahmood, I.~Rodriguez, and P.~Mogensen, ``Beyond 5g wireless irt for industry 4.0: Design principles and spectrum aspects,'' in \emph{2018 IEEE Globecom Workshops (GC Wkshps)}.\hskip 1em plus 0.5em minus 0.4em\relax IEEE, 2018, pp. 1--6.

\bibitem{aijaz2020private}
A.~Aijaz, ``Private 5g: The future of industrial wireless,'' \emph{IEEE Industrial Electronics Magazine}, vol.~14, no.~4, pp. 136--145, 2020.

\bibitem{ali2021urllc}
R.~Ali, Y.~B. Zikria, A.~K. Bashir, S.~Garg, and H.~S. Kim, ``Urllc for 5g and beyond: Requirements, enabling incumbent technologies and network intelligence,'' \emph{IEEE Access}, vol.~9, pp. 67\,064--67\,095, 2021.

\bibitem{gengtian2020power}
S.~Gengtian, T.~Koshimizu, M.~Saito, P.~Zhenni, L.~Jiang, and S.~Shimamoto, ``Power control based on multi-agent deep q network for d2d communication,'' in \emph{2020 International Conference on Artificial Intelligence in Information and Communication (ICAIIC)}.\hskip 1em plus 0.5em minus 0.4em\relax IEEE, 2020, pp. 257--261.

\bibitem{she2019ultra}
C.~She, C.~Liu, T.~Q. Quek, C.~Yang, and Y.~Li, ``Ultra-reliable and low-latency communications in unmanned aerial vehicle communication systems,'' \emph{IEEE Transactions on communications}, vol.~67, no.~5, pp. 3768--3781, 2019.

\end{thebibliography}

\end{document}